\documentclass[aps, prd, twocolumn, showpacs, showkeys, preprintnumbers, bibnotes, floatfix, longbibliography, notitlepage, nofootinbib, superscriptaddress, superscriptgaddress]{revtex4-2}

\usepackage{amsmath,amssymb,bm}
\usepackage[colorlinks=true, breaklinks=true]{hyperref}
\usepackage{graphicx,placeins}
\usepackage[italicdiff]{physics}
\usepackage{comment}
\usepackage{booktabs}
\usepackage{xcolor}
\usepackage{slashed}
\usepackage[dvipsnames]{xcolor}
\definecolor{darkred}{rgb}{0.5,0,0}
\definecolor{darkblue}{rgb}{0,0,0.5}
\definecolor{firebrick}{rgb}{0.75,0.125,0.125}
\definecolor{darkgreen}{rgb}{0,0.5,0}
\hypersetup{urlcolor=darkblue,
        citecolor=darkgreen,
        linkcolor=firebrick}
\usepackage{orcidlink}

\newcommand{\be}{\begin{equation}}
\newcommand{\ee}{\end{equation}}

\newcommand{\ie}{{\it i.e.}}

\newcommand{\eg}{{\it e.g.}}

\newcommand{\eq}{Eq.}

\newcommand{\fig}{Fig.}

\newcommand{\Refe}{Ref.}
\newcommand{\Refes}{Refs.}
\newcommand{\equ}[1]{\eq~(\ref{equ:#1})}
\newcommand{\figu}[1]{\fig~\ref{fig:#1}}

\makeatletter
\newcommand{\twocolclearpage}{%
  \close@column@grid
  \clearpage
  \twocolumngrid
}
\makeatother

\graphicspath{{figures/}}

\begin{document}

\title{Electron stability constrains neutrino time delays}

\author{Mauricio Bustamante}
\email{mbustamante@nbi.ku.dk}
\affiliation{Niels Bohr International Academy, Niels Bohr Institute,\\University of Copenhagen, 2100 Copenhagen, Denmark}

\author{Jos\'e Manuel Carmona}
\email{jcarmona@unizar.es}
\affiliation{Departamento de Física Teórica and Centro de Astropartículas y Física de Altas Energías (CAPA), Universidad de Zaragoza, Zaragoza 50009, Spain}

\author{Jos\'e Luis Cortés}
\email{cortes@unizar.es}
\affiliation{Departamento de Física Teórica and Centro de Astropartículas y Física de Altas Energías (CAPA), Universidad de Zaragoza, Zaragoza 50009, Spain}

\author{Ardit Gkioni}
\email{agkioni@unizar.es}
\affiliation{Departamento de Física Teórica and Centro de Astropartículas y Física de Altas Energías (CAPA), Universidad de Zaragoza, Zaragoza 50009, Spain}

\author{Maykoll A. Reyes}
\email{mkreyes@unizar.es}
\affiliation{Departamento de Física Teórica and Centro de Astropartículas y Física de Altas Energías (CAPA), Universidad de Zaragoza, Zaragoza 50009, Spain}

% \date{\today}
\date{July 1, 2026}

\begin{abstract}
Superluminal neutrino propagation, induced by Lorentz-invariance violation (LIV), is strongly constrained by vacuum pair emission, $\nu \to \nu + e^- + e^+$, a process ordinarily forbidden, which rapidly degrades the energy of high-energy neutrinos. Consequently, observable neutrino time delays are often preferentially associated with subluminal propagation, prompting LIV interpretations of claimed time delays between high-energy cosmic neutrinos and gamma rays. However, this expectation is at odds with the observed stability of high-energy electrons. The same Lorentz-violating correction associated with subluminal neutrino propagation opens the overlooked complementary decay channel $e^- \to e^- + \nu + \bar{\nu}$, leading to electron instability. We derive constraints on LIV from recent observations of TeV--PeV astrophysical electrons. These electron stability limits rule out LIV invoked to explain delays of high-energy cosmic neutrinos. Consequently, neutrino time delays are constrained on both the superluminal and subluminal sides. Therefore, observable delays require either purely astrophysical origins, a realization of LIV that affects all particle species equally, or physics beyond the standard effective-field-theory framework.
\end{abstract}

\maketitle

%%%%%%%%%%%%%%%%%%%%%%%%%%%%%%%%%%%%%%%%%%%%%%%%%%%%%%%%
%%%%%%%%%%%%%%%%%%%%%%%%%%%%%%%%%%%%%%%%%%%%%%%%%%%%%%%%

\textbf{Introduction.---}The discovery of Lorentz-invariance violation (LIV), a possible signature of quantum gravity, would disrupt modern physics~\cite{Amelino-Camelia:2008aez, Mattingly:2005re, Addazi:2021xuf, AlvesBatista:2023wqm}. Among the different possible manifestations of LIV, modified neutrino propagation stands out: because the standard propagation and interaction of neutrinos are well established, LIV-induced modifications may be readily observable~\cite{Kostelecky:2003cr, Kostelecky:2008ts, Kostelecky:2011gq}. We have identified an overlooked conflict between these modifications and the observed stability of high-energy electrons, putting the consistency of such LIV frameworks, and the interpretation of prior experimental claims, into question.

Reference~\cite{Cohen:2011hx} first showed that, under LIV, high-energy superluminal neutrinos traveling in vacuum may emit electron-positron pairs,
\begin{equation}
 \nu \to \nu + e^-+ e^+ \;,
 \label{equ:vpe}
\end{equation}
a process that is forbidden in the Standard Model. This \textit{vacuum pair emission} (VPE) renders superluminal neutrinos unstable via rapid energy degradation. This conflicts with the observation of TeV--PeV cosmic neutrinos, severely constraining superluminal LIV effects.

At the same time, phenomenological studies have explored the possibility that high-energy cosmic neutrinos tentatively associated to gamma-ray bursts exhibit time delays relative to the photons from them~\cite{ANTARES:2016fmg, Amelino-Camelia:2016ohi, Huang:2018ham, Ellis:2018ogq, Huang:2019etr, Amelino-Camelia:2022pja, Amelino-Camelia:2025lqn}. Since superluminal neutrino propagation is constrained by VPE, a subluminal LIV-induced slow-down is often regarded as the more viable possibility for observable delays.  (Recent analyses of the temporal structure of high-energy neutrino flares have begun to constrain this regime~\cite{Bustamante:2024fbj}.)

%%%%%%%%%%%%%%%%%%%%%%%%%%%%%%%%%%%%%%%%%%%%%%%%%%%%%%%%
\begin{figure}[t!]
 \centering
 \includegraphics[width=\columnwidth]{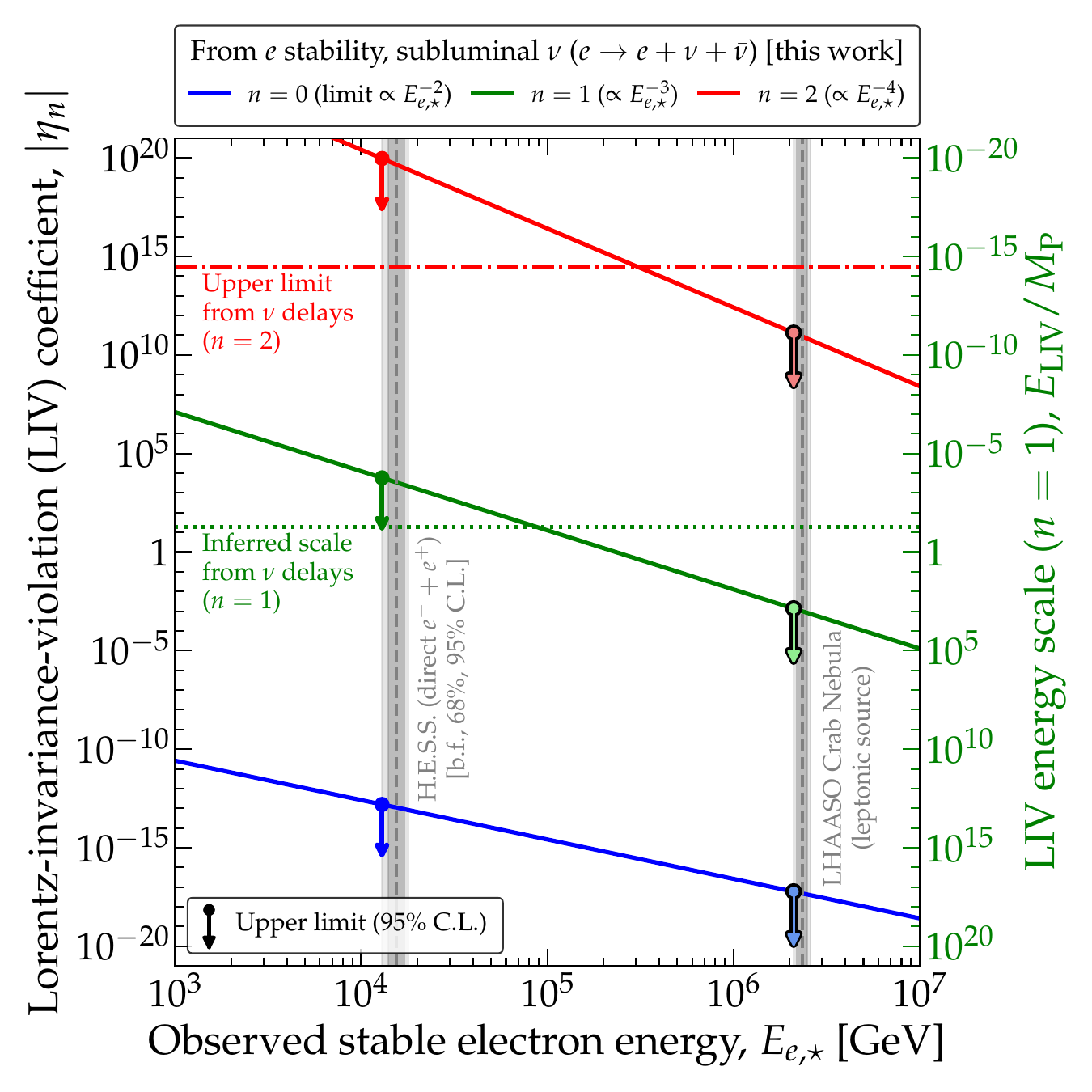}
 \vspace{-23pt}
 \caption{\textbf{Neutrino Lorentz-invariance violation limits from electron stability.} The survival of high-energy electrons imposes strict upper bounds on the Lorentz-violating coefficient $|\eta_n|$ to prevent the catastrophic electron decay $e^- \to e^- + \nu + \bar{\nu}$. Electron stability limits are shown from the 15.5-TeV electron-positron candidates observed by H.E.S.S.~\cite{HESS:2024etj} and the 1.12-PeV gamma ray observed by LHAASO (implying 2.34-PeV electrons)~\cite{LHAASO:2021cbz, LHAASO:2021gok}. The latter invalidates, by orders of magnitude, the subluminal-LIV parameter space previously invoked to explain anomalous neutrino-photon time delays with linear ($n = 1$) energy dependence~\cite{Amelino-Camelia:2016ohi, Huang:2018ham, Huang:2019etr, Amelino-Camelia:2022pja, Amelino-Camelia:2025lqn}, and establishes stringent new bounds on subluminal LIV effects with energy-independent ($n=0$), linear ($n=1$), and quadratic ($n=2$) neutrino-velocity modifications.}
 \label{fig:limits}
 \vspace*{-0.7cm}
\end{figure}
%%%%%%%%%%%%%%%%%%%%%%%%%%%%%%%%%%%%%%%%%%%%%%%%%%%%%%%%

In this Letter, we argue that this apparent asymmetry between superluminal and subluminal LIV-modified neutrino propagation is misleading. So far, it has been overlooked that under subluminal LIV modifications the electron can become unstable through
\begin{equation}
 e^-\to e^- + \nu+ \bar\nu \;.
 \label{equ:e_decay}
\end{equation}
We compute the decay width and energy loss for this process, constrain them using the observation of electron stability, and compare the constraints from VPE. 

%%%%%%%%%%%%%%%%%%%%%%%%%%%%%%%%%%%%%%%%%%%%%%%%%%%%%%%%
\begin{figure*}[t!]
 \centering
 \includegraphics[width=\textwidth]{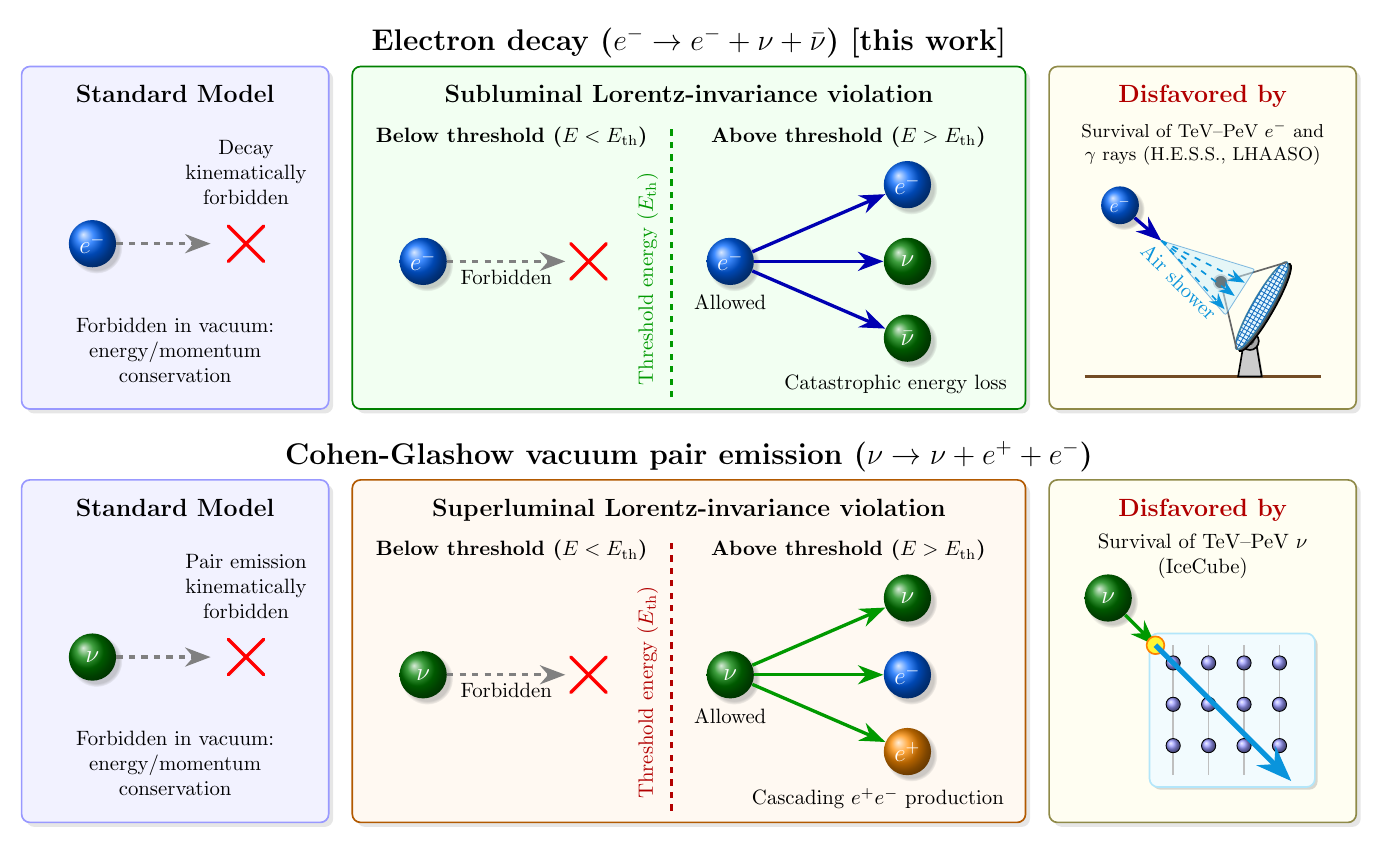}
 \vspace*{-0.7cm}
 \caption{\textbf{Anomalous vacuum decay processes induced by Lorentz-invariance violation (LIV).} While forbidden by exact energy-momentum conservation in the Standard Model (\textit{left panels}), LIV modifies particle dispersion relations, opening new decay and pair-emission channels above a critical energy threshold, $E_{\rm th}$ (\textit{center panels}). \textit{Top row:} The subluminal LIV-induced electron decay ($e^- \to e^- + \nu + \bar{\nu}$) derived and constrained in this work, which triggers catastrophic energy loss for high-energy electrons. \textit{Bottom row:} Contrast with the analogous superluminal Cohen-Glashow vacuum pair emission ($\nu \to \nu + e^+ + e^-$). \textit{Right panels:} depict the phenomenological survival arguments used to constrain these models: the detection of TeV--PeV cosmic-ray electrons via extensive air showers (\eg, in H.E.S.S., LHAASO) or cosmic neutrinos in neutrino telescopes (in IceCube) precludes such rapid energy losses, thereby placing stringent bounds on LIV.}
 \vspace*{-0.5cm}
 \label{fig:schematic}
\end{figure*}
%%%%%%%%%%%%%%%%%%%%%%%%%%%%%%%%%%%%%%%%%%%%%%%%%%%%%%%%

Figure~\ref{fig:limits} shows our results: electron stability challenges existing interpretations of time-delay claims of high-energy cosmic neutrinos as signatures of LIV.  We derive our limits for the LIV-induced electron decay by demanding the absence of catastrophic electron energy loss. Pitting our predictions against the observed survival of multi-TeV cosmic-ray electron candidates by H.E.S.S.~\cite{HESS:2024etj} and the inference of PeV-scale electrons in the Crab Nebula by LHAASO---based on the detection of PeV gamma rays~\cite{LHAASO:2021cbz, LHAASO:2021gok}---caps the value of the LIV coefficient that governs the modifications to neutrino propagation. These stability limits are compared to the values invoked in the literature to explain or constrain apparent time delays of TeV--PeV cosmic neutrinos relative to gamma rays from the same astrophysical sources.

Our stability limits derived from PeV electrons exclude these time-delay models. \textbf{\textit{Coupled to prior constraints on superluminal propagation, our results challenge the LIV interpretation of claimed time delays of high-energy cosmic neutrinos.}} Our analysis assumes LIV that affects only neutrinos, commonly explored in the literature. However, as we discuss below, our stability constraints are independent of this assumption. Evading these constraints would require a universal LIV deformation across all particle species---a scenario that, for modifications scaling linearly with energy, is unrealizable within standard effective field theory.

\medskip

%%%%%%%%%%%%%%%%%%%%%%%%%%%%%%%%%%%%%%%%%%%%%%%%%%%%%%%%
%%%%%%%%%%%%%%%%%%%%%%%%%%%%%%%%%%%%%%%%%%%%%%%%%%%%%%%%

\textbf{Neutrino LIV.---}We consider an isotropic LIV correction to neutrinos such that, in the ultrarelativistic limit, their dispersion relation is modified into
\begin{equation}
 E_\nu
 \simeq
 p+
 \frac{\eta_n}{2}
 \frac{p^{n+1}}{M_{\rm P}^n} \;,
 \label{eq:MDRnuMain}
\end{equation}
where $M_{\rm P}\simeq1.22\times10^{19}\,{\rm GeV}$ is the Planck mass and $\eta_n$ is a dimensionless constant representing the strength of LIV, which we assume affects all neutrino flavors equally. Our sign convention is such that $\eta_n>0$ corresponds to superluminal neutrino propagation. The correction is isotropic because it depends only on the magnitude of the particle momentum, $p$, not on its direction (anisotropic bounds would instead be line-of-sight dependent; for recent constraints, see, \eg, \Refes~\cite{Telalovic:2023tcb, Telalovic:2025xor}). Within an effective-field-theory LIV framework~\cite{Colladay:1998fq, Kostelecky:2003cr, Kostelecky:2011gq}, the corresponding antineutrino dispersion relation is~\cite{Carmona:2022dtp}
\begin{equation}
 E_{\bar\nu}
 \simeq
 p+
 (-1)^n
 \frac{\eta_n}{2}
 \frac{p^{n+1}}{M_{\rm P}^n} \;.
 \label{eq:MDRantinuMain}
\end{equation}
Therefore, for even $n$, neutrinos and antineutrinos receive corrections with the same sign, while for odd $n$ the signs are opposite. We consider $n=0,1,2$ as possible dominant behaviors for the neutrino velocity modification: an energy-independent shift and the leading linear and quadratic energy-dependent corrections. These modified dispersion relations make processes that are forbidden in the Standard Model kinematically allowed for particles with sufficiently high energy.  

Figure~\ref{fig:schematic} sketches the two LIV-induced processes we examine. For superluminal VPE, \equ{vpe}, the original estimate of the rate given in \Refe~\cite{Cohen:2011hx} was refined with a calculation of neutral- and charged-current contributions~\cite{Bezrukov:2011qn, Carmona:2012tp, Carmona:2022dtp}. 
For subluminal neutrino propagation, VPE is absent, but the complementary electron decay channel ($e{\rm D}$), \equ{e_decay}, can become kinematically allowed. For even $n$, this decay applies when $\eta_n<0$, so that both neutrinos and antineutrinos are subluminal. For odd $n$, changing the sign of $\eta_n$ exchanges the roles of neutrino and antineutrino; the decay is triggered by the subluminal state, and its kinematics depends on $|\eta_n|$. (Sans LIV, neutrinos are subluminal because they have mass, but their mass-induced delay at TeV--PeV is $\lesssim 10^{-9}$~s, negligible compared to the LIV delays discussed here.)

For both processes, the threshold energy, $E_{\rm th}$, above which they can occur takes the form
\begin{equation}
 E_{\rm th}^{n+2} =
 \begin{cases}
  4 \dfrac{m_e^2M_{\rm P}^n}{\eta_n} & ({\rm VPE},\ \eta_n>0) \\[2ex]
  C_n^{e{\rm D}} \dfrac{m_e^2M_{\rm P}^n}{|\eta_n|} & ( e{\rm D})
 \end{cases} \;.
 \label{equ:thresholds}
\end{equation}
Equation (\ref{equ:thresholds}) and the coefficients $C_n^{e{\rm D}}$ in it are derived in the Supplemental Material.  The smaller $|\eta_n|$ is, the higher the energy needed for the LIV effects to manifest.  Later, we turn this around to use the observation of high-energy neutrinos and electrons to bound the value of $\eta_n$.

\medskip

%%%%%%%%%%%%%%%%%%%%%%%%%%%%%%%%%%%%%%%%%%%%%%%%%%%%%%%%
%%%%%%%%%%%%%%%%%%%%%%%%%%%%%%%%%%%%%%%%%%%%%%%%%%%%%%%%

\textbf{Energy degradation.---}Sufficiently above threshold to neglect the electron mass~\cite{Carmona:2025lir}, the total decay widths for both processes take the generic asymptotic form
\begin{equation}
 \Gamma_n(E)
 =
 K_n
 \frac{G_F^2}{192\pi^3}
 E^{5+3n}
 \left(
 \frac{|\eta_n|}{M_{\rm P}^n}
 \right)^3 \;,
 \label{equ:genericWidthMain}
\end{equation}
where the numerical parameters $K_n$ are different for VPE and $e{\rm D}$. The values of $K_n$, as well as of $\kappa_n$ used below, are derived (for electron decay) and summarized (for VPE) in the Supplemental Material. 

The corresponding continuous energy-loss equation is
\begin{equation}
-\frac{dE}{dt}
=
\kappa_n E\Gamma_n(E),
\label{equ:genericLossMain}
\end{equation}
where $\kappa_n$ is the average fractional energy loss per decay. Defining $p_n=5+3n$, integration of \equ{genericLossMain} gives
\begin{equation}
 E_f^{-p_n}-E_i^{-p_n}
 =
 p_n\kappa_nK_n
 \frac{G_F^2}{192\pi^3}
 \left(
 \frac{|\eta_n|}{M_{\rm P}^n}
 \right)^3
 L \;.
 \label{eq:integratedLossMain}
\end{equation}
Due to the steep energy dependence of the loss rate, the final energy $E_f$ after propagation over a distance $L$ approaches the terminal value
\begin{equation}
 E_T^{(n)}
 =
 \left[
 p_n\kappa_nK_n
 \frac{G_F^2}{192\pi^3}
 \left(
 \frac{|\eta_n|}{M_{\rm P}^n}
 \right)^3
 L
 \right]^{-1/p_n} \;,
 \label{equ:terminalEnergyMain}
\end{equation}
provided that the initial energy, $E_i$, is above this energy scale, allowing us to frame our presentation below in terms of observed particle energies.

The asymptotic width in \equ{genericWidthMain} is reliable only sufficiently far above threshold. For this reason, we introduce a scaling factor $\rho>1$ and define $L^{(n)}$ as the distance for which $E_T^{(n)}=\rho E_{\rm th}$. Writing $E_{\rm th}^{n+2} = C_n m_e^2M_{\rm P}^n/|\eta_n|$, we obtain this energy-degradation length from \equ{terminalEnergyMain} as
\begin{equation}
 L^{(n)}
 =
 \frac{192\pi^3}
 {
 p_n\kappa_nK_nG_F^2
 (C_nm_e^2)^3
 }
 (\rho E_{\rm th})\,
 \rho^{-(p_n+1)} \;.
 \label{eq:LrhoMain}
\end{equation}
In the numerical estimates below we take $\rho=10$ as a conservative baseline, so that the terminal energy used to define $L^{(n)}$ lies well above threshold, in the regime where the asymptotic decay width is applicable. If $L^{(n)}$ is much shorter than the distance travelled by the neutrinos, or than the electron source size, particles injected above $\rho E_{\rm th}$ efficiently degrade to this energy before completing their propagation or escaping the source. Below, we use the energy-degradation length to assess the viability of superluminal and subluminal neutrino LIV.

\medskip

\textbf{Superluminal constraints.---}The condition $E_\nu^{\rm obs}<\rho E_{\rm th}^{\rm VPE}$ yields the superluminal bound $\eta_n < \eta_{n,{\rm sup}}^{\rm th} \equiv 4 m_e^2 M_{\rm P}^n (\rho / E_\nu^{\rm obs})^{n+2}$, where, expanded,
\begin{equation}
 \eta_{n,{\rm sup}}^{\rm th} \simeq \alpha_{n,{\rm sup}} \, \rho_{10}^{n+2} \left(\frac{E_\nu^{\rm obs}}{100\,{\rm PeV}}\right)^{-(n+2)} \;,
 \label{eq:etaSupNumMain}
\end{equation}
$\rho_{10} \equiv \rho/10$, and $\alpha_{n,{\rm sup}} \in \{1.0\times 10^{-20},\, 1.3\times 10^{-8},\, 1.6\times 10^{4}\}$ for $n=0, 1, 2$. The corresponding energy-degradation lengths, obtained from Eq.~\eqref{eq:LrhoMain}, are
\begin{equation}
 \frac{L_{\rm VPE}^{(n)}}{\text{pc}} \simeq \lambda_{n,{\rm sup}} \, \left( \frac{\rho E_{\rm th}^{\rm VPE}}{100\,{\rm PeV}} \right) \rho_{10}^{-(3n+6)} \;,
\end{equation}
with $\lambda_{n,{\rm sup}} \in \{30,\, 1.1\times 10^{-2},\, 7.1\times 10^{-6}\}$.
For $\rho=10$ and $\rho E_{\rm th}^{\rm VPE}=100\,{\rm PeV}$, these distances are orders-of-magnitude shorter than the nominal extragalactic, Gpc-scale baseline for high-energy cosmic neutrinos. Thus, a neutrino produced above $\rho E_{\rm th}^{\rm VPE}$ would degrade to this energy before reaching Earth; consequently, observing a neutrino with energy $E_\nu^{\rm obs}$ implies $\eta_n < \eta_{n,{\rm sup}}^{\rm th}$. Therefore, the KM3-230213A event detected by the KM3NeT neutrino telescope~\cite{KM3NeT:2025npi}, if interpreted as a cosmic neutrino with energy of order $100\,{\rm PeV}$, automatically implies the superluminal LIV bound in Eq.~\eqref{eq:etaSupNumMain}~\cite{KM3NeT:2025mfl, Carmona:2025lir}.

\medskip

%%%%%%%%%%%%%%%%%%%%%%%%%%%%%%%%%%%%%%%%%%%%%%%%%%%%%%%%
%%%%%%%%%%%%%%%%%%%%%%%%%%%%%%%%%%%%%%%%%%%%%%%%%%%%%%%%

\textbf{Subluminal constraints.---}Let $E_{e,\star}$ denote the highest electron energy that is directly observed or inferred from gamma-ray observations of a leptonic astrophysical source. The condition $E_{e,\star}<\rho E_{\rm th}^{e{\rm D}}$ yields the subluminal bound $|\eta_n| < \eta_{n,{\rm sub}}^{\rm th} \equiv C_n^{e{\rm D}} m_e^2 M_{\rm P}^n (\rho / E_{e,\star})^{n+2}$, where
\begin{equation}
 \eta_{n,{\rm sub}}^{\rm th} \simeq \alpha_{n, {\rm sub}} \, \rho_{10}^{n+2} \left(\frac{E_{e,\star}}{100\,{\rm TeV}}\right)^{-(n+2)} \;,
 \label{equ:eDetaNumMain}
\end{equation}
with $\alpha_{n, {\rm sub}} \in \{2.6\times10^{-15},\, 1.3\times10^{1},\, 2.6\times10^{16}\}$ for $n=0, 1, 2$.
The relevant comparison scale is now not a Gpc baseline, but the residence, cooling, or acceleration scale in the source environment. Using Eq.~\eqref{eq:LrhoMain}, one finds
\begin{equation}
 \frac{L_{e{\rm D}}^{(n)}}{\text{pc}} \simeq \lambda_{n,{\rm sub}}^{\rm th} \, \left( \frac{\rho E_{\rm th}^{e{\rm D}}}{100\,{\rm TeV}} \right) \rho_{10}^{-(3n+6)} \;,
\label{eq:LrhoEDMain}
\end{equation}
with $\lambda_{n,{\rm sub}}^{\rm th} \in \{1.3,\, 4.7\times10^{-4},\, 1.6\times10^{-8}\}$.
For $n=1,2$, the energy-degradation lengths are negligible on cosmic scales in the multi-TeV--PeV range. For $n=0$, the length is of parsec size when $E_{e,\star}=\rho E_{\rm th}^{e{\rm D}} =100\,{\rm TeV}$, so the  interpretation is more dependent on the source environment. Nevertheless, this value is still well below the typical extension of pulsar TeV halos, which can reach tens of parsecs~\cite{Martin:2024cpo}. In all cases, the bound strengthens rapidly with the maximum electron energy, $|\eta_n|_{\rm max}\propto E_{e,\star}^{-(n+2)}$. 

Direct space-based measurements of cosmic-ray electrons place $E_{e,\star}$ in the multi-TeV range~\cite{DAMPE:2017fbg, CALET:2023emo}, while ground-based H.E.S.S. observations extend the electron-positron candidates to tens of TeV~\cite{HESS:2024etj}. Interpreting the Crab Nebula as a leptonic source, the connection between MeV synchrotron emission and multi-TeV inverse-Compton gamma rays implies the existence of electrons above 1~PeV~\cite{HEGRA:2004tpc}, as do the flares observed by \textit{Fermi}-LAT~\cite{Fermi-LAT:2010pdh}, and gamma rays seen by HAWC beyond 100~TeV~\cite{HAWC:2019xhp} and by LHAASO at PeV~\cite{LHAASO:2021cbz, Shi:2026qtx}. 

\medskip

%%%%%%%%%%%%%%%%%%%%%%%%%%%%%%%%%%%%%%%%%%%%%%%%%%%%%%%%
%%%%%%%%%%%%%%%%%%%%%%%%%%%%%%%%%%%%%%%%%%%%%%%%%%%%%%%%

\textbf{Neutrino time delays.---}Neutrinos propagating over a distance $L_\nu$ accrue an LIV time shift of magnitude
\begin{equation}
 |\Delta t_n|
 \simeq
 \frac{n+1}{2}
 |\eta_n|
 \left(
 \frac{E_\nu}{M_{\rm P}}
 \right)^n
 L_\nu \;.
 \label{eq:timedelayMain}
\end{equation}
The superluminal bounds on $\eta_n$ constrain delays to
\begin{equation}
 \frac{|\Delta t_n|_{\rm sup}}{\text{s}} \lesssim \tau_{n,{\rm sup}} \, \left(\frac{L_\nu}{1\,{\rm Gpc}}\right) \rho_{10}^{n+2} \left(\frac{E_\nu^{\rm obs}}{100\,{\rm PeV}}\right)^{-2} ,
\label{eq:TDSupMain}
\end{equation}
with $\tau_{n,{\rm sup}} \in \{5.4\times10^{-4},\, 1.1\times10^{-2},\, 1.6\times10^{-1}\}$ for $n=0, 1, 2$. Thus, superluminal LIV compatible with the observation of a 100-PeV neutrino by KM3NeT~\cite{KM3NeT:2025npi} cannot generate delays longer than subsecond-scale over a Gpc baseline.

For subluminal propagation, electron stability gives
\begin{align}
 \frac{|\Delta t_n|_{\rm sub}}{\text{s}} &\lesssim \tau_{n,{\rm sub}} \, \left(\frac{L_\nu}{1\,{\rm Gpc}}\right) \rho_{10}^{n+2} \left(\frac{E_\nu}{1\,{\rm PeV}}\right)^n \nonumber \\
 & \quad \times~\left(\frac{E_{e,\star}}{100\,{\rm TeV}}\right)^{-(n+2)} \;,
 \label{eq:TDSubMain}
\end{align}
with $\tau_{n,{\rm sub}} \in \{1.3\times10^{2},\, 1.1\times10^{5},\, 2.7\times10^{7}\}$. For PeV cosmic neutrinos like those observed by IceCube~\cite{IceCube:2013cdw}, the stability of 100-TeV electrons makes year-long neutrino delays, such as those discussed in \Refe~\cite{Amelino-Camelia:2025lqn}, borderline admissible, and only in the optimistic $n=2$ case; 1-PeV electrons shrink this to 45 minutes. 

\medskip

\textbf{Contrasting LIV bounds.---}Figure~\ref{fig:limits} shows our stability limits on subluminal $|\eta_n|$ derived using \equ{eDetaNumMain} from observations by H.E.S.S.~($E_{e,\star} \!=\! 15.5 \pm 1.55$~TeV~\cite{HESS:2024etj}) and LHAASO ($E_{e,\star} = 2.34 \pm 0.145$~PeV, inferred from a photon with $1.12 \pm 0.09$~PeV~\cite{LHAASO:2021cbz}). We estimate the 95\% confidence level (C.L.) lower value on $E_{e,\star}$ from a one-sided Gaussian and use it to report upper limits at 95\% C.L.: $|\eta_n| \lesssim 1.5 \times 10^{-13}$, $6.0 \times 10^3$, $9.3 \times 10^{19}$ for $n = 0$, 1, 2 from H.E.S.S.; and $5.8 \times 10^{-18}$, $1.4 \times 10^{-3}$, $1.3 \times 10^{11}$ from LHAASO.

To contextualize our limits, we compare them to the parameter spaces conventionally invoked to explain anomalous time delays of high-energy cosmic neutrinos. Prior studies have explored candidate associations between high-energy IceCube neutrinos and gamma-ray flares, where the neutrinos arrive with an energy-dependent delay relative to the photons. Assuming these delays stem from modified propagation, these studies measure or constrain the LIV scale, $E_{\text{LIV}, n}$.

For linear ($n=1$) LIV, an analysis of delays in multiple IceCube events claimed a measurement of the LIV energy scale of $E_{\text{LIV},1} = (6.5 \pm 0.4) \times 10^{17}$~GeV~\cite{Huang:2018ham}. For  quadratic ($n=2$) LIV, explaining these delays imply a lower energy scale due to the steeper energy dependence. Instead of a measurement, \Refe~\cite{Wang:2016lne} used the maximum possible travel delay from a 2-PeV IceCube event associated with a flare from the blazar PKS~B1424-418 to constrain this quadratic scale to a lower limit of $E_{\text{LIV},2} \gtrsim 7.3 \times 10^{11}$~GeV. Expressed in terms of our dimensionless coefficient $|\eta_n| = (M_{\rm P} / E_{\text{LIV},n})^n$, these delay-based results correspond to $|\eta_1| \approx 19$ and $|\eta_2| \lesssim 2.8 \times 10^{14}$. Separately, \Refe~\cite{Amelino-Camelia:2016fuh} reported $|\eta_1| \approx 22 \pm 2$. As shown in \figu{limits}, these claimed subluminal scales fall orders of magnitude deep into the excluded regime based on LHAASO PeV gamma-ray measurements, indicating that such time delays cannot be generated by generic subluminal kinematics without catastrophically destroying PeV electrons.

\medskip

%%%%%%%%%%%%%%%%%%%%%%%%%%%%%%%%%%%%%%%%%%%%%%%%%%%%%%%%
%%%%%%%%%%%%%%%%%%%%%%%%%%%%%%%%%%%%%%%%%%%%%%%%%%%%%%%%

\textbf{Summary and outlook.---}Lorentz-invariance violation (LIV) inducing superluminal neutrino propagation is strictly constrained by vacuum pair emission, $\nu \to \nu + e^- + e^+$. We have shown that this is only one facet of a broader stability problem: subluminal neutrino LIV, which slows down rather than speeds up neutrinos, inevitably opens the complementary electron decay channel, $e^-\to e^- + \nu + \bar{\nu}$. Driven by decay rates that scale steeply with energy and trigger catastrophic energy losses, electron stability provides a stringent, overlooked bound on subluminal LIV.

This instability creates a severe tension for neutrino time-delay models. Superluminal LIV compatible with the survival of an observed 100-PeV cosmic neutrino generates subsecond delays over Gpc baselines. Subluminal delays are conversely bounded by electron stability. For LIV modifications linear in energy ($n=1$), generating the years-long delays claimed for TeV--PeV neutrinos associated with gamma-ray bursts requires an LIV coefficient so large that electrons above $100\,{\rm TeV}$ would rapidly decay. Similarly, for quadratic modifications ($n=2$), accommodating the months-long delays inferred from blazar flares demands coefficients thoroughly invalidated by PeV electron observations. Because astrophysics confirms electron survival up to $1\,{\rm PeV}$, these delay-invoked parameter spaces are robustly ruled out. Consequently, \textbf{\textit{long neutrino time delays cannot be attributed to generic subluminal LIV, countering prior claims.}}

Relaxing the assumption that LIV affects only neutrinos does not bypass this conclusion. Introducing LIV in the electron sector would merely shift the constraints: closing the electron-decay channel requires the electron to be at least as subluminal as the neutrino, re-opening the complementary neutrino decay channel. This logic extends universally; differing LIV coefficients between electrons, photons (triggering photon decay~\cite{LIVQEDInPrep}), and ultra-high-energy protons trigger analogous energy losses---in the latter case, even more pronounced ones, given the energies reached are higher.

Avoiding all instabilities would require a universal subluminal LIV deformation common to neutrinos, charged leptons, and hadrons. However, a universal linear scenario is unrealizable within standard effective field theory (EFT), where neutrinos and antineutrinos receive LIV corrections of opposite signs. Consequently, observable neutrino time delays are difficult to accommodate in generic LIV EFTs. Their observation---barring purely astrophysical explanations---would evidence not just a modified neutrino velocity, but a universally shared or non-EFT quantum structure of spacetime.

%%%%%%%%%%%%%%%%%%%%%%%%%%%%%%%%%%%%%%%%%%%%%%%%%%%%%%%%
%%%%%%%%%%%%%%%%%%%%%%%%%%%%%%%%%%%%%%%%%%%%%%%%%%%%%%%%

\medskip

\textbf{Acknowledgments.---}MB is supported by {\sc Villum Fonden} under project no.~29388.  The authors acknowledge support by the Spanish grant PID2024-160228NB-I00, funded by the Agencia Estatal de Investigación (MCIN/AEI/10.13039/501100011033), and by the Gobierno de Aragón through the research group E21\_23R (Grupo Teórico de Física de Altas Energías). AG is supported by the predoctoral contract CPE\_27\_25 funded by the Gobierno de Aragón under the 2025–2029 programme for predoctoral research personnel. The authors would like to acknowledge the contribution of the COST Actions CA18108 ``Quantum gravity phenomenology in the multi-messenger approach'' and CA23130 ``Bridging high and low energies in search of quantum gravity''. MB is grateful to CAPA and Departamento de Física Teórica at the University of Zaragoza for hosting him during the conception of this work.

%%%%%%%%%%%%%%%%%%%%%%%%%%%%%%%%%%%%%%%%%%%%%%%%%%%%%%%%
%%%%%%%%%%%%%%%%%%%%%%%%%%%%%%%%%%%%%%%%%%%%%%%%%%%%%%%%

% \bibliography{refs}

%apsrev4-2.bst 2019-01-14 (MD) hand-edited version of apsrev4-1.bst
%Control: key (0)
%Control: author (8) initials jnrlst
%Control: editor formatted (1) identically to author
%Control: production of article title (0) allowed
%Control: page (0) single
%Control: year (1) truncated
%Control: production of eprint (0) enabled
%

\clearpage
\appendix

%%%%%%%%%%%%%%%%%%%%%%%%%%%%%%%%%%%%%%%%%%%%%%%%%%%%%%%%
%%%%%%%%%%%%%%%%%%%%%%%%%%%%%%%%%%%%%%%%%%%%%%%%%%%%%%%%

\onecolumngrid

\begin{center}
 \large
 Supplemental Material for\\
 \smallskip
 \textit{Electron stability constrains neutrino time delays}
\end{center}

\twocolumngrid

%%%%%%%%%%%%%%%%%%%%%%%%%%%%%%%%%%%%%%%%%%%%%%%%%%%%%%%%
%%%%%%%%%%%%%%%%%%%%%%%%%%%%%%%%%%%%%%%%%%%%%%%%%%%%%%%%

\section{Kinematics of electron decay}
\label{app:kinematics}

We derive the threshold conditions for
\begin{equation}
 e^-(p)\to e^-(p')+\nu(k)+\bar\nu(q) \;.
\label{eq:smProcess}
\end{equation}
The initial electron is assumed to obey the standard relativistic dispersion relation, while the neutrino and antineutrino satisfy
\begin{align}
 E_\nu
 &\simeq
 |\vec k|
 +
 \frac{\eta_n}{2}
 \frac{|\vec k|^{n+1}}{M_{\rm P}^n} \;,
 \\
 E_{\bar\nu}
 &\simeq
 |\vec q|
 +
 (-1)^n
 \frac{\eta_n}{2}
 \frac{|\vec q|^{n+1}}{M_{\rm P}^n} \;.
\end{align}
Taking the initial three-momentum $\vec p$ as the longitudinal direction, we write
\begin{equation}
 \vec p\,{}'
 =
 x\vec p+\vec p\,{}'_T \;,
 \quad
 \vec k
 =
 y\vec p+\vec k_T \;,
 \quad
 \vec q
 =
 z\vec p-\vec p\,{}'_T-\vec k_T \;,
\end{equation}
with
\begin{align}
 & x+y+z=1 \;,
 \quad
 0<x,y,z<1 \;,
 \label{equ:sum_rule}
 \\
 & \vec p\,{}'_T\cdot\vec p=\vec k_T\cdot\vec p=0 \;.
\end{align}
The collinear expansions of the particle energies are:
\begin{align}
 E_e(\vec p\,)
 &\simeq
 |\vec p\,|
 +
 \frac{m_e^2}{2|\vec p\,|} \;,
 \\
 E_e(\vec p\,{}')
 &\simeq
 x|\vec p\,|
 +
 \frac{m_e^2+|\vec p\,{}'_T|^2}{2x|\vec p\,|} \;,
 \\
 E_\nu(\vec k)
 &\simeq
 y|\vec p\,|
 +
 \frac{|\vec k_T|^2}{2y|\vec p\,|}
 +
 \frac{\eta_n}{2}
 \frac{y^{n+1}|\vec p\,|^{n+1}}{M_{\rm P}^n} \;,
 \\
 E_{\bar\nu}(\vec q)
 &\simeq
 z|\vec p\,|
 +
 \frac{|\vec p\,{}'_T+\vec k_T|^2}{2z|\vec p\,|}
 \nonumber \\
 & \quad +
 (-1)^n
 \frac{\eta_n}{2}
 \frac{z^{n+1}|\vec p\,|^{n+1}}{M_{\rm P}^n} \;.
\end{align}

From here, energy conservation gives
\begin{align}
 &
 \frac{|\vec p\,{}'_T|^2}{x}
 +
 \frac{|\vec k_T|^2}{y}
 +
 \frac{|\vec p\,{}'_T+\vec k_T|^2}{z}
 \\
 &\hspace{1cm}
 =
 -\eta_n
 \frac{|\vec p\,|^{n+2}}{M_{\rm P}^n}
 \left[
 y^{n+1}
 +
 (-1)^n z^{n+1}
 \right]
 -
 m_e^2
 \frac{1-x}{x} \;.
 \nonumber
 \label{eq:smElectronBalance}
\end{align}
Since the left-hand side is non-negative, the decay is allowed only if the first term on the right-hand side, due to LIV, can compensate the second term. For the signs of $\eta_n$ for which electron decay is possible, this condition can be written as
\begin{equation}
 |\eta_n|
 \frac{|\vec p\,|^{n+2}}{M_{\rm P}^n}
 \,\mathcal F_n(y,z)
 \ge
 m_e^2
 \frac{1-x}{x} \;,
 \label{eq:smGeneralCondition}
\end{equation}
where
\begin{equation}
 \mathcal F_n(y,z)=
 \begin{cases}
  y^{n+1}+z^{n+1},
  & n\ {\rm even},\ \eta_n<0,
  \\[1ex]
  |y^{n+1}-z^{n+1}|,
  & n\ {\rm odd}.
 \end{cases}
 \label{equ:smFnDefinition}
\end{equation}
For even $n$, this corresponds to both the neutrino and antineutrino being subluminal. For odd $n$, only one of the two is subluminal, and the absolute value in \equ{smFnDefinition} selects the configuration in which that particle carries the larger longitudinal fraction.

For fixed $x$, the maximum of $\mathcal F_n$ with $y+z=1-x$ [\equ{sum_rule}] is, for $n\ge1$,
\begin{equation}
 \max_{y+z=1-x}\mathcal F_n(y,z)=(1-x)^{n+1} \;,
\end{equation}
and is reached at the boundary where one of the two neutrinos is soft. Therefore, the threshold energy for the decay to occur is obtained from
\begin{equation}
 |\eta_n|
 \frac{|\vec p\,|^{n+2}}{M_{\rm P}^n}
 x(1-x)^n
 \ge
 m_e^2 \;.
\end{equation}
Maximizing $x(1-x)^n$ over $0<x<1$ gives
\begin{equation}
 \max_x x(1-x)^n
 =
 \frac{n^n}{(n+1)^{n+1}} \;,
\end{equation}
and, hence,
\begin{equation}
 E_{\rm th}^{n+2}
 =
 C_n^{e{\rm D}}
 \frac{m_e^2M_{\rm P}^n}{|\eta_n|} 
 ~~(\text{for}~n \geq 1) \;,
 \label{equ:threshold_energy}
\end{equation}
where the constant
\begin{equation}
 C_n^{e{\rm D}}
 =
 \frac{(n+1)^{n+1}}{n^n} 
 ~~(\text{for}~n \geq 1) \;.
 \label{equ:smGeneralThreshold}
\end{equation}

The case $n=0$ is special. In that case, $\mathcal F_0=y+z=1-x$, and the threshold energy, \equ{threshold_energy}, reduces to
\begin{equation}
 E_{\rm th}^2
 =
 \frac{m_e^2}{|\eta_0|} \;.
\end{equation}
For the coefficients $C_n^{e{\rm D}}$, we use \equ{smGeneralThreshold} for $n \geq 1$ and 
\begin{equation}
 C_0^{e{\rm D}}=1 \;,
\end{equation}
for $n = 0$.

A similar threshold analysis for VPE gives
\begin{equation}
 E_{\rm th}^{\,n+2}=
 \frac{4m_e^2M_{\rm P}^n}{\eta_n} \;.
\end{equation}

%%%%%%%%%%%%%%%%%%%%%%%%%%%%%%%%%%%%%%%%%%%%%%%%%%%%%%%%
%%%%%%%%%%%%%%%%%%%%%%%%%%%%%%%%%%%%%%%%%%%%%%%%%%%%%%%%

\section{Decay width}

We use the notation of Ref.~\cite{Carmona:2022dtp} for external neutrino states in the presence of LIV. The spinors associated with the neutrino and antineutrino external legs are denoted by $\tilde u$ and $\tilde v$, respectively. They are solutions of the LIV-modified free Dirac equation. In our convention,
\begin{align}
 \left[
 \gamma^0 E_\nu
 -
 \vec\gamma\cdot\vec k
 -
 \gamma^0
 \frac{\eta_n}{2}
 \frac{E_\nu^{n+1}}{M_{\rm P}^n}
 \right]\tilde u(k)
 &=0 \;,
 \\
 \left[
 \gamma^0 E_{\bar\nu}
 -
 \vec\gamma\cdot\vec q
 +
 (-1)^{n+1}
 \gamma^0
 \frac{\eta_n}{2}
 \frac{E_{\bar\nu}^{n+1}}{M_{\rm P}^n}
 \right]\tilde v(q)
 &=0 \; .
\end{align}
To leading order in $E_\nu/M_{\rm P}$ and $E_{\bar{\nu}}/M_{\rm P}$, this implies
\begin{align}
 E_\nu
 &\simeq
 |\vec k|
 +
 \frac{\eta_n}{2}
 \frac{|\vec k|^{n+1}}{M_{\rm P}^n} \;,
 \\
 E_{\bar\nu}
 &\simeq
 |\vec q|
 +
 (-1)^n
 \frac{\eta_n}{2}
 \frac{|\vec q|^{n+1}}{M_{\rm P}^n} \;.
\end{align}
We distinguish the physical energies, $E_\nu$ and $E_{\bar\nu}$, which satisfy energy conservation, from the zeroth components of momenta entering the leading spin sums. Defining the auxiliary lightlike four-vectors
\begin{equation}
 \hat k^\mu=(|\vec k|,\vec k) \;,
 \qquad
 \hat q^\mu=(|\vec q|,\vec q) \;,
\end{equation}
we have, in the collinear approximation,
\begin{equation}
 \tilde u(k)\,\bar{\tilde u}(k)
 =
 P_L\,\slashed{\hat k} \;,
 \quad
 \tilde v(q)\,\bar{\tilde v}(q)
 =
 P_L\,\slashed{\hat q} \;,
 \label{eq:tildeSpinSums}
\end{equation}
where $P_L=(1-\gamma_5)/2$ is the left-handed projector. Thus, the weak matrix element can be evaluated with the usual ultrarelativistic traces, while the LIV dependence enters the decay rate through the modified energy conservation and the resulting phase-space factor. In the following, scalar products involving neutrino momenta in the squared amplitude are understood to be evaluated with the auxiliary lightlike four-vectors, $\hat k,\hat q$; hats are omitted to simplify the notation.

The neutral-current contribution to
\begin{equation}
 e^-(p)\to e^-(p')+\nu_\beta(k)+\bar\nu_\beta(q)
\end{equation}
is mediated by $Z^0$ exchange. For a left-handed final electron and $\beta\neq e$, the amplitude of this process is
\begin{align}
 \mathcal M_L^{(\beta\neq e)}
 &=
 i\,\frac{g^2}{2M_W^2}
 \left(s_W^2-\frac12\right)
 \nonumber \\
 & \quad \times \left[\bar{\tilde u}(k)\gamma_\mu\tilde v(q)\right]
 \left[\bar u_L(p')\gamma^\mu u_L(p)\right] \;,
\end{align}
where $g$ is the weak isospin gauge coupling constant, $s_W  \equiv \sin\theta_W$, and $\theta_W$ is the Weinberg angle. (We neglect the electron mass in the calculation; this is a good approximation when the process occurs sufficiently above threshold.) For a right-handed final electron,
\begin{equation}
\mathcal M_R
=
i\,\frac{g^2}{2M_W^2}
s_W^2
\left[\bar{\tilde u}(k)\gamma_\mu\tilde v(q)\right]
\left[\bar u_R(p')\gamma^\mu u_R(p)\right].
\end{equation}
For $\beta=e$, there is an extra charged-current contribution,
\begin{equation}
\mathcal M_W
=
i\,\frac{g^2}{2M_W^2}
\left[\bar{\tilde u}(k)\gamma_\mu u_L(p)\right]
\left[\bar u_L(p')\gamma^\mu\tilde v(q)\right] \;.
\end{equation}
The spin-summed squared amplitudes are
\begin{align}
 |\mathcal M_L^{(\beta\neq e)}|^2
 &=
 4\left(\frac{g^2}{M_W^2}\right)^2
 \left(s_W^2-\frac12\right)^2 
 \nonumber \\
 & \quad \times (p\cdot q)(p'\cdot k) \;, \\
 %%%%
 |\mathcal M_L^{(\beta=e)}|^2
 &=
 4\left(\frac{g^2}{M_W^2}\right)^2
 \left(s_W^2-\frac32\right)^2
 \nonumber \\
 & \quad \times (p\cdot q)(p'\cdot k) \;, \\
 %%%%
 |\mathcal M_R|^2
 &=
 4\left(\frac{g^2}{M_W^2}\right)^2
 s_W^4
 (p\cdot k)(p'\cdot q) \;,
\end{align}
Averaging over the initial electron spin gives
\begin{align}
 \overline{|\mathcal M|^2}
 &=
 2\left(\frac{g^2}{M_W^2}\right)^2
 \left[
 C_{L,\beta}^2(p\cdot q)(p'\cdot k)
 \right.
 \nonumber \\
 & \quad \left. +~
 s_W^4(p\cdot k)(p'\cdot q)
 \right] \;,
\label{eq:smMsq}
\end{align}
where
\begin{equation}
 C_{L,\beta}
 =
 \begin{cases}
  s_W^2-\dfrac12,
  & \beta=\mu,\tau,
  \\[0.3cm]
  s_W^2-\dfrac32,
  & \beta=e.
 \end{cases}
 \label{equ:CLbeta}
\end{equation}

Following the collinear phase-space integration developed in \Refe~\cite{Carmona:2022dtp}, the decay width can be written as an integral over the longitudinal momentum fractions $x$, $y$, and $z$ of the final-state electron, neutrino, and antineutrino, respectively (Appendix~\ref{app:kinematics}). In the process we consider, the initial electron has no LIV modification, and so the relevant LIV factor is
\begin{equation}
 \Delta_n(y,z)
 =
 -\frac{\eta_n}{2}
 \left[
 y^{n+1}
 +
 (-1)^n z^{n+1}
 \right] \;.
\label{eq:smDelta}
\end{equation}
The decay is possible only when $\Delta_n>0$. The partial width is
\begin{align}
 \Gamma_{\beta,n}^{e{\rm D}}(E)
 &\simeq
 \mathcal{N}
 \left(\frac{E}{M_{\rm P}}\right)^{3n}
 \int dx\,dy\,dz\,
 \delta(1-x-y-z)
 \nonumber \\
 & \quad \times \Theta[\Delta_n(y,z)]
 \Delta_n^3(y,z)
 \nonumber \\
 & \quad \times \left[
 C_{L,\beta}^2(1-z)^2
 +
 s_W^4(1-y)^2
 \right] \;,
 \label{eq:smWidthIntegral}
\end{align}
where
\begin{equation}
 \mathcal N
 \equiv
 \frac12
 \left(\frac{g^2}{M_W^2}\right)^2
 \frac{E^5}{192\pi^3} \;.
 \label{equ:rate_prefactor}
\end{equation}

For $n=1$ and $\eta_1>0$,
\begin{equation}
 \Delta_1(y,z)=\frac{\eta_1}{2}(z^2-y^2) \;,
\end{equation}
and the condition $\Delta_1>0$ gives $z>y$. With $z=1-x-y$, the integration region is
\begin{equation}
 0<x<1,
 \qquad
 0<y<\frac{1-x}{2}.
\end{equation}
Thus,
\begin{align}
 \Gamma_{\beta,1}^{e{\rm D}}(E)
 &\simeq
 \mathcal N
 \left(\frac{E}{M_{\rm P}}\right)^3
 \eta_1^3
 \int_0^1 dx
 \int_0^{(1-x)/2}dy\,
 \frac{(z^2-y^2)^3}{8}
 \nonumber\\
 &\quad\times
 \left[
 C_{L,\beta}^2(1-z)^2
 +
 s_W^4(1-y)^2
 \right].
\end{align}
Performing the $y$ integration yields
\begin{align}
 \Gamma_{\beta,1}^{e{\rm D}}(E)
 &\simeq
 \mathcal N
 \left(\frac{E}{M_{\rm P}}\right)^3
 \eta_1^3
 \int_0^1 dx\,
 \frac{(1-x)^7}{3840}
 \\
 &\quad\times
 \left[
 C_{L,\beta}^2(49x^2+10x+1)
 +
 s_W^4(x^2+10x+49)
 \right] \nonumber \;.
\end{align}
The final result is
\begin{equation}
 \Gamma_{\beta,1}^{e{\rm D}}(E)
 \simeq
 \mathcal N
 \left(\frac{E}{M_{\rm P}}\right)^3
 \eta_1^3
 \left[
 \frac{C_{L,\beta}^2}{9600}
 +
 \frac{47s_W^4}{28800}
 \right].
 \label{eq:smWidthN1}
\end{equation}

For $n=0$ and $\eta_0<0$,
\begin{equation}
 \Delta_0(y,z)
 =
 \frac{|\eta_0|}{2}(y+z)
 =
 \frac{|\eta_0|}{2}(1-x) \;, 
\end{equation}
which is positive in the full region, $y>0$, $z>0$, $y+z<1$. Thus,
\begin{align}
 \Gamma_{\beta,0}^{e{\rm D}}(E)
 &\simeq
 \mathcal N|\eta_0|^3
 \int_0^1 dx
 \int_0^{1-x}dy\,
 \frac{(1-x)^3}{8}
 \nonumber\\
 &\quad\times
 \left[
 C_{L,\beta}^2(x+y)^2
 +
 s_W^4(1-y)^2
 \right] \;.
\end{align}
Performing the $y$ integration yields
\begin{equation}
 \Gamma_{\beta,0}^{e{\rm D}}(E)
 \simeq
 \mathcal N|\eta_0|^3
 \int_0^1 dx\,
 \frac{(1-x)^3(1-x^3)}{24}
 \left[
 C_{L,\beta}^2+s_W^4
 \right],
 \label{equ:decay_width_n_0_y_integrated}
\end{equation}
and, therefore,
\begin{equation}
 \Gamma_{\beta,0}^{e{\rm D}}(E)
 \simeq
 \mathcal N|\eta_0|^3
 \frac{17}{1680}
 \left[
 C_{L,\beta}^2+s_W^4
 \right] \;.
 \label{eq:smWidthN0}
\end{equation}

For $n=2$ and $\eta_2<0$,
\begin{equation}
 \Delta_2(y,z)
 =
 \frac{|\eta_2|}{2}(y^3+z^3) \;.
\end{equation}
Thus,
\begin{align}
 \Gamma_{\beta,2}^{e{\rm D}}(E)
 &\simeq
 \mathcal N
 \left(\frac{E}{M_{\rm P}}\right)^6
 |\eta_2|^3
 \int_0^1 dx
 \int_0^{1-x}dy\,
 \frac{(y^3+z^3)^3}{8}
 \nonumber\\
 &\quad\times
 \left[
 C_{L,\beta}^2(1-z)^2
 +
 s_W^4(1-y)^2
 \right].
\end{align}
Performing the $y$ integration yields
\begin{align}
 \Gamma_{\beta,2}^{e{\rm D}}(E)
 &\simeq
 \mathcal N
 \left(\frac{E}{M_{\rm P}}\right)^6
 |\eta_2|^3
 \int_0^1 dx\,
 \frac{(1-x)^{10}}{8}
 \label{equ:decay_width_n_1_y_integrated}
 \\
 &\quad\times
 \left[
 \frac{29}{140}x
 +
 \frac{73}{840}(1-x)^2
 \right]
 \left[
 C_{L,\beta}^2+s_W^4
 \right] \nonumber \;,
\end{align}
and, therefore,
\begin{equation}
 \Gamma_{\beta,2}^{e{\rm D}}(E)
 \!\simeq\!
 \mathcal N
 \left(\frac{E}{M_{\rm P}}\right)^6
 |\eta_2|^3
 \frac{661}{640640}
 \left[
 C_{L,\beta}^2+s_W^4
 \right] \;.
 \label{equ:decay_width_n_2_y_integrated}
\end{equation}

Setting $\sin^2\theta_W \approx 1/4$, $C_{L, \beta}$ in \equ{CLbeta} evaluate to
\begin{equation}
 C_{L,\mu,\tau}^2=\frac{1}{16},
 \qquad
 C_{L,e}^2=\frac{25}{16} \;.
\end{equation}
Writing the prefactor $\mathcal{N}$ [\equ{rate_prefactor}] in terms of the Fermi constant, $G_F \equiv \sqrt{2} g^2 / (8 M_W^2)$, the general expression for the decay width into $\nu_\beta$ and $\bar{\nu}_\beta$ becomes
\begin{equation}
 \Gamma_{\beta,n}^{e{\rm D}}(E)
 =
 K_{\beta,n}^{e{\rm D}}
 \frac{G_F^2}{192\pi^3}
 E^{5+3n}
 \left(
 \frac{|\eta_n|}{M_{\rm P}^n}
 \right)^3 \;,
\end{equation}
where the coefficients are
\begin{align}
 K_{\mu,\tau,0}^{e{\rm D}}
 &=
 \frac{17}{840} \;,
 &
 K_{e,0}^{e{\rm D}}
 &=
 \frac{221}{840} \;,
 \\[1ex]
 K_{\mu,\tau,1}^{e{\rm D}}
 &=
 \frac{1}{576} \;,
 &
 K_{e,1}^{e{\rm D}}
 &=
 \frac{61}{14400} \;,
 \\[1ex]
 K_{\mu,\tau,2}^{e{\rm D}}
 &=
 \frac{661}{320320} \;,
 &
 K_{e,2}^{e{\rm D}}
 &=
 \frac{8593}{320320} \;.
\end{align}

The total decay width, summing over all flavors, is $\Gamma_{n}^{e{\rm D}} = \sum_{\beta = e, \mu, \tau} \Gamma_{\beta,n}^{e{\rm D}}$, \ie,
\begin{equation}
 \Gamma_{n}^{e{\rm D}}(E)
 =
 K_{n}^{e{\rm D}}
 \frac{G_F^2}{192\pi^3}
 E^{5+3n}
 \left(
 \frac{|\eta_n|}{M_{\rm P}^n}
 \right)^3 \;,
\end{equation}
where $K_n^{e{\rm D}} = K_{e,n}^{e{\rm D}} + 2K_{\mu,\tau,n}^{e{\rm D}}$, which, explicitly, is
\begin{equation}
 K_0^{e{\rm D}}=\frac{17}{56} \;,
 \quad
 K_1^{e{\rm D}}=\frac{37}{4800} \;,
 \quad
 K_2^{e{\rm D}}=\frac{1983}{64064} \;.
\end{equation}

%%%%%%%%%%%%%%%%%%%%%%%%%%%%%%%%%%%%%%%%%%%%%%%%%%%%%%%%
%%%%%%%%%%%%%%%%%%%%%%%%%%%%%%%%%%%%%%%%%%%%%%%%%%%%%%%%

\section{Mean energy loss}

In the LIV-induced decay, $e \to e + \nu + \bar{\nu}$,  let $x$ be the fraction of the initial electron energy carried by the final electron. The average energy fraction retained by the electron in the decay is
\begin{equation}
\langle x\rangle_n^{e{\rm D}}
=
\frac{
\displaystyle
\sum_\beta
\int_0^1 dx\,
x\,
\frac{d\Gamma_{\beta,n}^{e{\rm D}}}{dx}
}{
\displaystyle
\sum_\beta
\Gamma_{\beta,n}^{e{\rm D}}
} \;,
\end{equation}
where the differential decay width, $d\Gamma_{\beta, n}^{e{\rm D}}/dx$, is calculated from Eqs.~(\ref{equ:decay_width_n_0_y_integrated}), (\ref{equ:decay_width_n_1_y_integrated}), and (\ref{equ:decay_width_n_2_y_integrated}) for $n = 0, 1, 2$.

The fractional energy loss per decay is
\begin{equation}
 \kappa^{\rm eD}_n=1-\langle x\rangle^{\rm eD}_n \;.
\end{equation}
Evaluated, this yields
\begin{equation}
 \begin{split}
  \kappa_0^{e{\rm D}}&=\frac{55}{68}\simeq 0.809,
  \qquad
  \kappa_1^{e{\rm D}}=\frac{697}{814}\simeq 0.856, \\
  \quad
  \kappa_2^{e{\rm D}}&=\frac{4224}{4627}\simeq 0.913.
 \end{split}
 \label{eq:eDkappaMain}
\end{equation}
Thus, a single decay removes most of the electron energy. 

\section{VPE coefficients}

Here we summarize the coefficients $K_n$ in \equ{genericWidthMain} in the main text, and the  fractional energy loss for the VPE process used in this work, taken from \Refes~\cite{Carmona:2022dtp,Carmona:2025lir}. Averaging over flavors after oscillations,
\begin{equation}
 K_0^{\rm VPE}=\frac{17}{84} \;,
 \qquad
 K_1^{\rm VPE}=\frac{121}{336} \;,
 \qquad
 K_2^{\rm VPE}=\frac{81}{182} \;.
 \label{eq:VPEKMain}
\end{equation}
The corresponding mean fractional energy losses are
\begin{equation}
 \begin{split}
  \kappa_0^{\rm VPE} &= \frac{55}{68} \simeq 0.809 \;,
  \qquad
  \kappa_1^{\rm VPE} = \frac{2966}{3993} \simeq 0.743 \;, \\
  \kappa_2^{\rm VPE} & = \frac{31}{44} \simeq 0.705 \;.
 \end{split}
 \label{eq:VPEkappaMain}
\end{equation}

\end{document}